\documentclass[doublecol]{epl2} 
\usepackage{amsmath}
\usepackage{epsfig}
\usepackage{bm}
\usepackage{graphicx}
\usepackage{amsfonts}
\usepackage{amssymb}

\title{Geometric properties of two-dimensional coarsening 
with weak disorder}

\author{Alberto Sicilia\inst{1} \and Jeferson J. Arenzon\inst{2} \and Alan J. Bray\inst{3} \and Leticia F. Cugliandolo\inst{1} }

\institute{
  \inst{1} Universit\'e Pierre et Marie Curie -- Paris VI, LPTHE UMR 7589, - 
4 Place Jussieu,  75252 Paris Cedex 05, France\\
  \inst{2} Instituto de F\'\i sica, Universidade Federal do 
Rio Grande do Sul -  CP 15051, 91501-970 Porto Alegre RS, Brazil\\
  \inst{3} School of Physics and Astronomy, University of Manchester - Manchester M13 9PL, UK
}

\abstract{The domain morphology of weakly disordered ferromagnets,  
quenched from the high-temperature phase to the low-temperature phase, 
is studied using numerical simulations. We find that the geometrical 
properties of the coarsening domain structure, e.g.\ the distributions 
of hull enclosed areas and domain perimeter lengths, are described 
by a scaling phenomenology in which the growing domain scale 
$R(t)$ is the only relevant parameter. Furthermore, the scaling functions 
have forms identical to those of the corresponding pure system, extending 
the ``super-universality'' property previously noted for the pair correlation 
function.
}

\begin{document}

\maketitle

\section{Introduction}

Many aspects of the out of equilibrium relaxation of macroscopic
systems still remain to be unveiled. The mechanism underlying
coarsening phenomena, or the domain growth of two competing
equilibrium phases after a quench from the disordered phase, is well
understood~\cite{BrayReview}.  However, an important part of the
description of this process is phenomenological and, to a certain
extent, qualitative.

A common feature of nearly all coarsening systems is that their
dynamics are described by a dynamical scaling hypothesis with a single
characteristic growing length scale $R(t)$ that depends on the 
problem at hand. This hypothesis states that 
the domain morphology is statistically the same at all times
when lengths are measured in units of $R(t)$. As a result, 
all time-dependences in correlation functions are encoded in $R$ and
lengths appear scaled by this typical length. The scaling hypothesis 
has been well-verified experimentally, numerically as well as 
analytically in a small number of solvable simple cases. It is 
thus a well-established feature of phase ordering dynamics 
though no generic proof exists.  

Most studies of coarsening dynamics focus on the growth law $R(t)$ and
the scaling functions of various correlators and linear responses.  An
analytic determination of these functions remains open -- apart from 
some toy models such as the one-dimensional ferromagnetic chain or the 
$O(N)$ model in the large $N$ limit. A number of field-theoretic
approaches to finding an approximate form of the scaling 
functions of two-point and two-time correlations have been
proposed but none of them is fully successful~\cite{BrayReview}.

In~\cite{us-PRL,us-PRE} we analysed the coarsening process in pure
(not disordered) two-dimensional systems with non-conserved order parameter from a
geometric point of view. We studied the probability distributions of
hull-enclosed and domain areas and hull and domain-wall lengths.  By
deriving an analytic expression for the number density of
hull-enclosed areas we proved that the scaling hypothesis holds for
this quantity. With a few additional assumptions we extended these
results to the statistical properties of domains and boundary lengths and
we put them all to the numerical test.  In this Letter we extend these
results to the coarsening dynamics of two-dimensional models with
non-conserved order parameter under the effect of {\it weak quenched
disorder}. By `weak disorder' we mean randomness that does not modify the
character of the ordered phase (see~\cite{PuriReview} for a recent
review).

An important property of domain-growth in systems with weak disorder
is the super-universality hypothesis that states that once the correct
growing length scale is taken into account all scaling
functions are independent of the disorder strength.  This conjecture
was first enunciated by Fisher and Huse~\cite{FisherHuse} using a
renormalization-group approach. The idea is that the length scale at
which the effects of quenched disorder are important is much smaller
than the domain scale $R(t)$. The latter dominates the elastic
energy. Thus the dynamics of large structures is approximately 
curvature driven. Pinning at small scales modifies the scale
factor, that is to say the growth law $R(t)$, but not the scaling
functions.  The validity of super-universality for the scaling function of the
equal-time two-point correlation of several disordered systems
including the random bond Ising model was checked numerically
in~\cite{BrayHumayun,PuriChowdhury,Hayakawa,Iwai}. The stringest test
proposed in~\cite{Blundell} that includes higher order correlations
was also passed numerically in~\cite{Biswal}. 

In short, in this Letter we continue our geometric study
of coarsening in two-dimensional systems. We simulate the dynamics
of the random bond Ising model and we pay special attention to
the scaling and super-universality properties of the probability
distribution of areas and borders.

\section{Hull enclosed area distribution}
\label{sec:hull}

The ordering process of a magnetic system in its low temperature 
phase is visualized in terms of domains or
regions of connected aligned spins. Each domain has
one external perimeter which is called the {\it hull}. The {\it hull
enclosed area} is the total area within this perimeter. 
In~\cite{us-PRL,us-PRE} we derived an analytic expression for the hull
enclosed area distribution of pure, curvature driven, two-dimensional 
coarsening with
non-conserved order parameter. Using a continuum description in which
the non conserved order parameter is a scalar field we found that the
number of hull enclosed areas 
per unit area, $n^p_h(A,t)\,dA$, with enclosed area in
the interval $(A,A+dA)$, is
\begin{equation}
n^p_h(A,t)  =  \frac{2c_h}{(A+\lambda_h t)^2}
\; . 
\label{eq:analytic-nh}
\end{equation}
$c_h=1/8\pi\sqrt{3}$ is a universal constant that enters this
expression through the influence of the initial condition and was
computed by Cardy and Ziff in their study of the geometry of critical
structures in equilibrium~\cite{Cardy}.  $\lambda_h$ is a material
dependent constant relating the local velocity of an interface and its
local curvature in the Allen-Cahn equation, $v=-(\lambda_h/2\pi) \,
\kappa$~\cite{AC}. The expression (\ref{eq:analytic-nh}) is valid for
all quenches from equilibrium in the high temperature phase,
$T_0>T_c$, to zero working temperature, $T=0$. For a critical initial
condition, $T_0=T_c$, the same expression holds with $2c_h$ replaced
by $c_h$. Equation~(\ref{eq:analytic-nh}) can be recast in the scaling
form
\begin{equation}
n^p_h(A,t) = \frac{1}{(\lambda_h t)^2} \; \; 
f\left( \frac{A}{\lambda_h t} \right)
\label{eq:scaling}
\end{equation}
with $f(x)=\frac{2c_h}{(x+1)^2}$.  In this way, scaling with the
characteristic length scale, $R_p(t) = \sqrt{\lambda_h t}$, for
coarsening dynamics with non conserved order parameter in a pure
system is demonstrated. The effects of a finite working
temperature are fully encoded in the temperature dependence of the
parameter $\lambda_h$ while the same scaling function $f(x)$ describes
$n^p_h$~\cite{us-PRE} as suggested by the zero temperature fixed point
scenario~\cite{BrayReview}.

When quenched disorder is introduced the growing phenomenon is no
longer fully curvature driven and domain-wall pinning by disorder
becomes relevant. At early times, the system avoids pinning and
evolves like in the pure case. Later, barriers pin the domain walls
and the system gets trapped in metastable states from which it can
escape only by thermal activation over the corresponding free-energy
barriers.  In spite of these differences, coarsening in ferromagnetic
systems with quenched disorder also satisfies dynamic
scaling~\cite{BrayHumayun,PuriChowdhury,Hayakawa,Iwai,Biswal,Blundell}
and a single characteristic length, $R(t,T,\varepsilon)$, can be
identified ($\varepsilon$ is a measure of the disorder amplitude).
As a result of the competition between the curvature driven mechanism
and pinning by disorder, the coarsening process is slowed down and the
characteristic radius of the domains depends on the disorder strength 
and it is smaller than the pure one, $R(t,T,\varepsilon) <
R_p(t,T)$. Moreover, the super-universality hypothesis applied to the
scaling function of the equal-time two-point correlation function in
random ferromagnets was verified numerically in a number of
works~\cite{BrayHumayun,PuriChowdhury}.

The scaling and super-universality hypotheses suggest that
Eq.~(\ref{eq:scaling}) remains valid with the same scaling function
$f(x)=2c_h/(1+x)^2$ and $(\lambda_h t)^{1/2}$ 
replaced by $R(t,T,\varepsilon)$ for all $2d$
non-conserved order parameter coarsening processes in which the
low-temperature ordered phase is not modified.  More precisely,
we expect that 
\begin{equation}
n_h(A,t,T,\varepsilon) = R^{-4}(t,T,\varepsilon) \; \; f\left[ \frac{A}{R^2(t,T,\varepsilon)} \right]
\;
\label{eq:scaling2}
\end{equation}
should be valid in all these cases.
In this Letter we test this hypothesis by following the 
dynamic evolution of the 
two-dimensional random bond Ising model ($2d$ RBIM) defined by the
Hamiltonian,
\begin{equation}
H = - \sum_{\langle i,j\rangle} J_{ij} \sigma_i \sigma_j
\label{model}
\end{equation}
where the $J_{ij}$ are random variables uniformly distributed over the
interval $[1-\varepsilon/2, 1+\varepsilon/2]$ with $0 < \varepsilon
\leq 2$. This model has a second-order phase transition between a
high-temperature paramagnetic phase and a low-temperature
ferromagnetic phase.  We simulate the dynamic evolution of a model
defined on a square lattice with linear size $L=10^3$, using
a single-flip Monte Carlo technique and the heat-bath algorithm.
Data are averaged over $10^3$ samples. We show results for a random initial condition, 
$\sigma_i=\pm 1$ with equal probability to mimic an infinite temperature
equilibrium state, $T_0\to\infty$. At the initial time $t=0$ we set
the working temperature to a low value and we follow the evolution
thereafter.  We are interested on the effect of quenched randomness
and we thus focus on a single working temperature, $T=0.4$, at which
the ordered equilibrium phase is ferromagnetic for all $0\leq \varepsilon\leq
2$. In what follows we drop the $T$ dependence from $R$ and $n_h$, so we simply
denote them $R(t,\varepsilon)$ and $n_h(A,t,\varepsilon)$.

\begin{figure}[h]
\onefigure[width=8.5cm]{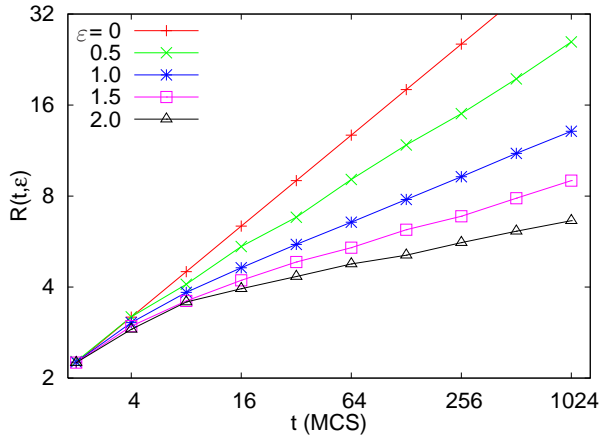}
\caption{(Colour online.) Log-log plot of the characteristic domain
  radius extracted from the collapse of the equal-time two-point
  correlation function against time after the quench. Different curves
  correspond to different values of the quenched disorder strengths. 
  For comparison we include the not disordered limit law
  $R(t,\varepsilon=0) \propto \sqrt{t}$.}
\label{fig:growing-law}
\end{figure}

We determine the growth law, $R(t,\varepsilon)$, 
from a direct measure of the
spatial correlation function $C(r,t,\varepsilon)$:
\begin{eqnarray}
C(r,t,\varepsilon) \equiv
\frac{1}{N} \sum_{i=1}^N 
\langle \, s_i(t) s_j(t) \, \rangle
\Large{|}_{|\vec r_i - \vec r_j|=r}
\end{eqnarray}
 which also obeys dynamical scaling,
\begin{eqnarray}
C(r,t,\varepsilon)
\sim  m^2(T,\varepsilon) \; f\left(\frac{r}{R(t,\varepsilon)}\right)
\; ,
\label{eq:Crt}
\end{eqnarray}
with $m(T,\varepsilon)$ the equilibrium magnetisation density.  In
Fig.~\ref{fig:growing-law} we show $R(t,\varepsilon)$ as a function of
$t$ for several values of the quenched disorder strength along with the
pure case. We extracted $R(t,\varepsilon)$ by collapsing the
equal-time correlations data using Eq.~\ref{eq:Crt}.
After a few time-steps all curves deviate from the pure
$R(t,\varepsilon=0)\propto \sqrt{t}$ law and the subsequent growth is
the slower the stronger the disorder strength $\varepsilon$. For
strong disorder the characteristic length reaches relatively modest
values during the simulation interval, for instance,
$R(t=1024,\varepsilon=2)\sim 7$ (measured in units of the lattice
spacing). It is then quite hard to determine the actual functional law
describing the late time evolution of $R(t,\varepsilon)$ and it comes as
no surprise that this issue has been the matter of debate
recently~\cite{HuseHenley,Oh,Rieger,Rieger2,Rieger3,Henkel1,Henkel2,Hinrichsen}.
Using arguments based on the energetics of domain-wall pinning Huse
and Henley~\cite{HuseHenley} proposed the law $R(t)\propto(\ln
t)^{1/\psi}$ with $\psi=1/4$ for the random bond Ising model.  More
recently, powerful Montecarlo
simulations~\cite{Oh,Rieger,Rieger2,Rieger3,Henkel1,Henkel2,Hinrichsen} suggest
a power law $R(t)\propto t^{\theta}$ with a exponent $\theta$ that
depends on $T$ and $\varepsilon$.  We are not concerned here with
predicting the time-dependence of $R$. Instead, we use the numerical
values in Fig.~\ref{fig:growing-law} to scale our data for the
distributions of hull enclosed areas and hull lengths as explained
below.

\begin{figure}[h]
\onefigure[width=8.5cm]{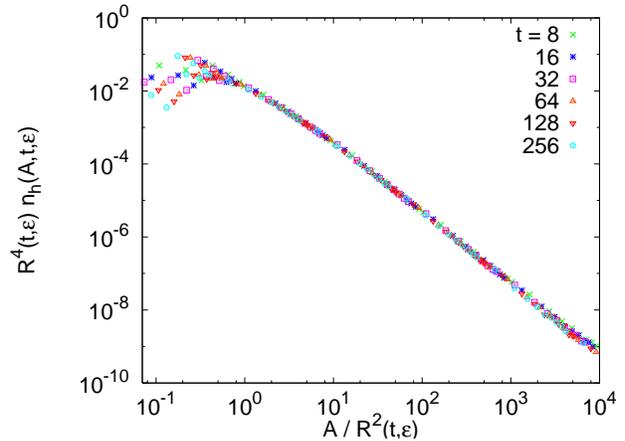}
\caption{(Colour online.) Number density of hull enclosed areas for
  one value of the disorder strength, $\varepsilon=2$, at several
  times. The data are shown in the scaling form
  and they collapse for $A/R^2 \stackrel{>}{\sim}1$.}
\label{fig:nh-scaling}
\end{figure}

In Fig.~\ref{fig:nh-scaling} we display data for one disorder
strength, $\varepsilon=2$, taken at several times after the
quench. Using the scaling form the data collapse for $A/R^2
\stackrel{>}{\sim}1$ over 8 decades in the vertical axis and 4 decades
in the horizontal axis. Deviations are seen for small areas and we
discuss their possible origin below.  This analysis confirms the
scaling hypothesis in the region of large areas, $A/R^2
\stackrel{>}{\sim}1$.

\begin{figure}
\onefigure[width=8.5cm]{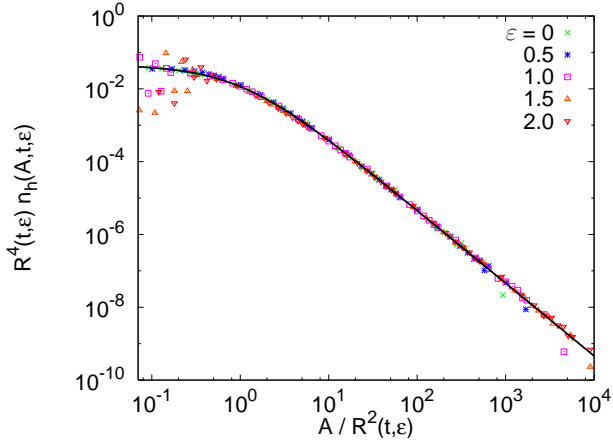}
\caption{(Colour online.) Number density of hull enclosed areas 
at $t=256$ MCs and several values of the disorder strength.
The collapse of all data for $A/R^2 \stackrel{>}{\sim}1$
supports the validity of super-universality. The black solid line represents 
the analytic result for the pure case.} 
\label{fig:nh-hyperscaling}
\end{figure}

In Fig.~\ref{fig:nh-hyperscaling} we test the super-universality hypothesis
as applied to the hull-enclosed area distribution by presenting data
for a single instant, $t=256$ MCs, and several values of the disorder
strength, $\varepsilon=0, \, 0.5, \, 1, \, 1.5, \, 2$. Again, for areas
such that $A/R^2 \stackrel{>}{\sim}1$ there is a very accurate data
collapse while for smaller areas deviations, that we discuss below,
are visible. The solid black line represents the analytic prediction
for the pure case that yields, under the super-universality hypothesis,
the analytic prediction for all $\varepsilon$.  The curve falls
well on top of the data. In the tail of the distribution we see
downward deviations that are due to finite size effects already
discussed in detail in \cite{us-PRE}.

Let us now examine several possible origins for the deviation of the
numerical data from the analytic prediction at small areas, $A/R^2
\stackrel{<}{\sim} 1$, in Figs.~\ref{fig:nh-scaling} and
\ref{fig:nh-hyperscaling}.  The first source of problems could be the
fact that we need to use a finite working temperature in the
disordered case to depin the walls while the analytic results are
derived at $T=0$.  In the pure case the effect of temperature is to
create thermal domains within the genuinely coarsening ones. Based on
this observation in \cite{us-PRE} we explained that the small area
probability distribution may have an excess contribution coming from
these thermal fluctuations.  By extracting the contribution of the
{\it equilibrium} distribution of thermal domains, that became
important for temperatures $T\stackrel{>}{\sim} 1$, we showed that the
number density of hull-enclosed areas at $T>0$ is given by the zero
temperature result once scaled by $R(T,t)$. In the present disordered
case, the working temperature we use is too low to generate any
thermal domains and thus temperature cannot be the source of deviations
from the analytic form.

Another possible origin of the difference between numerical data and
theoretical prediction is the fact that the analytic results are
derived using a continuum field-theoretic description of coarsening
while numerical simulations are done on a lattice.  In the presence of
quenched randomness one can expect the effects of the lattice
discretization to be more important than in the pure case, especially
for relatively small structures. Moreover, disorder induces domain-wall
roughening and the Allen-Cahn flat interface assumption is not as well 
justified.

Finally, the super-universality hypothesis is argued for large structures
only~\cite{FisherHuse} and thus the small hull enclosed areas are not
really forced to follow it strictly when $A\stackrel{<}{\sim} R^2$.

\section{Geometric structure of the domains}
\label{sec:structure}

\begin{figure}[h]

\onefigure[width=8.5cm]{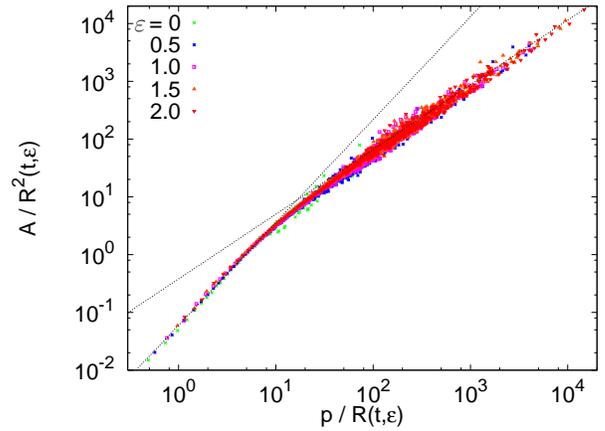}
\caption{(Colour online.) 
Averaged scatter plot of the scaled areas against perimeters 
for several values of the strength of the quenched randomness.}
\label{fig:perimetervssurface}
\end{figure}

In order to better characterise the geometric structure of the
coarsening process we study the relation between hull-enclosed areas
and perimeters.  In Fig.~\ref{fig:perimetervssurface} we present the
average of the scatter plot of the scaled hull enclosed area, $A/R^2$,
against the corresponding scaled perimeter, $p/R$, in a log-log plot
for several values of the disorder strength at $t=256$ MCs.  We
observe a separation into two regimes at the breaking point $p^*/R
\sim 10$ and $A^*/R^2\sim 2$. In the upper and lower regime areas and
perimeters are related by two power laws that {\it do not} depend on
the disorder strength.  More precisely~\cite{us-PRE},
\begin{equation}
\frac{A}{R^2(t,\varepsilon)} 
\sim 
\left[ \frac{p}{R(t,\varepsilon)} \right]^{\alpha}
\label{eq:perimetervssurface}
\end{equation}
with 
\begin{eqnarray} 
&& {\alpha}^> ={\alpha^>_0}=1.12\pm 0.10
  \;\;\;\;\;\;\mbox{for} \;\;\;\;\;\; \frac{p}{R(t,\varepsilon)} \;
  \stackrel{>}{\sim} 10 \; ,
\nonumber\\
&& {\alpha}^<={\alpha^<_0} \sim 1.83 \pm 0.10
  \;\;\;\;\;\;\mbox{for} \;\;\;\;\;\; \frac{p}{R(t,\varepsilon)} \;
  \stackrel{<}{\sim} 10 \; .
\label{eq:alpha}
\end{eqnarray}
The upper exponent also characterises the highly ramified structures
of the initial condition. Note that the small value of $\alpha^>$ is not 
related to the existence of holes in these large structures,
as suggested for domains in~\cite{Jacobs}, since 
hull-enclosed areas do not have holes in them. The crossover between upper and
lower regimes depends on time when observed in absolute value. In
other words, during the coarsening process a caracteristic scale
$p^{*}\propto R(t,\varepsilon)$ develops such that hull-enclosed areas
with perimeter $p>p^{*}$ have the same exponent $\alpha_0\sim 1.12$ as in
the initial condition before the quench~\cite{us-PRE} while
hull-enclosed areas with smaller perimeter are more compact as
indicated by the smaller value of the exponent $\alpha^<$.

\section{Hull length distribution}

\begin{figure}
\onefigure[width=8.5cm]{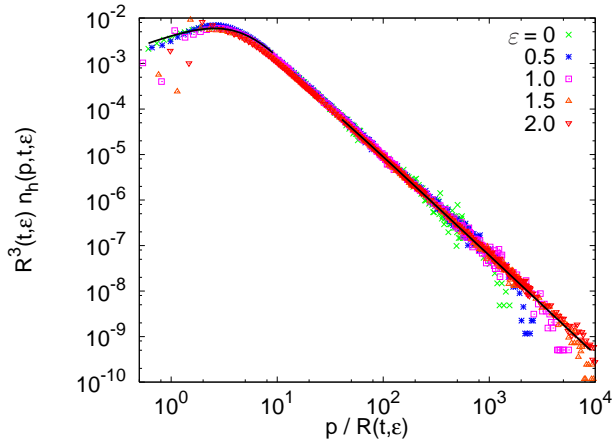}
\caption{(Colour online.) Number density of hull lengths at $t=256$
  MCs and several values of the disorder strength given in the
  key. The collapse of all data for $p/R \stackrel{>}{\sim} 2$
  supports the validity of super-universality in this regime of lengths.}
\label{fig:np-hyperscaling}
\end{figure}

We now examine the 
hull length distribution that is expected to
follow the scaling and super-universality hypotheses. In
Fig.~\ref{fig:np-hyperscaling} we display the number density of hull
lengths at $t=256$ MCs and $\varepsilon=0,\, 0.5,\, 1,\, 1.5, \, 2$.
We show the data in the form suggested by the scaling hypothesis 
\begin{equation}
 n_h(p,t,\varepsilon) = R^{-3}(t,\varepsilon) \; g\left[ \frac{p}{R(t,\varepsilon)} \right]
\; . 
\end{equation}
The data collapse on a master curve for all lengths satisfying $p/R
\stackrel{>}{\sim} 2$ a value that is roughly the location of the
maximum in the distribution. The super-universality hypothesis holds in this 
regime of lengths. 

Relating the distribution functions of areas and perimeters through a 
change of variables, as explained in \cite{us-PRE}, we find
\begin{equation}
R^3(t,\varepsilon)
\;
n_h(p,t,\varepsilon) \sim 
\frac{
\alpha^{\stackrel{>}{<}} c_h \left(
\frac{p}{R(t,\varepsilon)}\right)^{\alpha^{\stackrel{>}{<}}-1}} 
{\left[ 
1+ \left(\frac{p}{R(t,\varepsilon)}\right)^{\alpha^{\stackrel{>}{<}}} 
\right]^{2}
}
\; . 
\label{eq:np}
\end{equation}
The scaling function is thus the same as in the pure
system~\cite{us-PRE} with two branches characterized by the 
exponents $\alpha^>$ and $\alpha^<$. As in the pure case
the maximum in the distribution is described by 
Eq.~(\ref{eq:np}) with $\alpha^<$ while the maximum found with 
$\alpha^>$ falls outside its range of validity. 

\begin{figure}[h]
\onefigure[width=8.5cm]{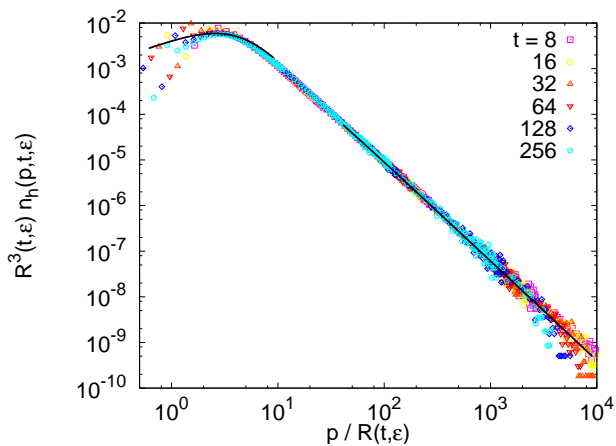}
\caption{(Colour online.) Scaling plot of the number density of hull lengths 
for a disorder with strength $\varepsilon=1.0$ at different times. The solid black lines are the analytic prediction
in Eq.~(\ref{eq:np}) and explained in the text. The part around the 
maximum is characterized by the exponent $\alpha^<$ while the tail of the 
distribution is well-described by the exponent $\alpha^>$.}
\label{fig:nh-perimeters}
\end{figure}

In Fig.~\ref{fig:nh-perimeters} we display the number density of hull
lengths scaled with the typical radius $R$ for one value of the
disorder strength at several times after the quench.  The upper and
lower predictions are shown with solid black lines. 

\section{Conclusions}

In this Letter we analysed the statistics of hull-enclosed areas and
hull lengths during the coarsening dynamics of the $2d$ RBIM with a 
uniform distribution of coupling strengths. We found
that the number densities of these observables satisfy scaling and
super-universality for structures with $A/R^2\stackrel{>}{\sim} 1$
and $p/R\stackrel{>}{\sim} 2$.

We showed that the analytic prediction for the number density of hull
enclosed areas derived for pure systems also describes the statistics
of these quantities in the presence of quenched ferromagnetic
disorder. The geometric properties of the boundaries between phases
are, in principle, more sentitive to quenched randomness than their interior. 
We showed, however, that the relation between areas and interfaces and, 
in consequence, the distribution of hull lengths are independent of 
the disorder strength also satisfying super-universality. 

In previous work all numerical tests of the super-universality hypothesis
in the RBIM have focused on the study the equal time two-point correlation
function~\cite{BrayHumayun,PuriChowdhury,Hayakawa,Iwai,Biswal}. 
The geometric approach that we have presented enables us to show that 
this property applies even to very small structures.

We also studied a number of related problems on which we   report below. 
\begin{itemize}

\item 
We verified that analogous results are obtained for 
different probability distributions of the coupling strengths
as long as these remain ferromagnetic. 

\item
We observed that the scaling plots do not depend on the working 
temperature while the latter is below the critical point. 
 
\item 
All our results can be easily extended to the study of domain 
areas and walls as done in \cite{us-PRE} for the pure case.

\item
We checked that the super-universality hypothesis is also verified in
$d=1$~\cite{d1,us-PRE} for relatively long domains. Small 
deviations for very short ones are also observed in this case.

\end{itemize}

\acknowledgments JJA, LFC and AS acknowledge financial support from
Capes-Cofecub research grant 448/04. JJA is partially supported by the
Brazilian agencies CNPq and FAPERGS. LFC is a member of Institut
Universitaire de France.

\end{document}